\documentclass[osajnl,preprint,showpacs,superscriptaddress,12pt]{revtex4-1} %% use 12pt for preprint option
\usepackage{amsmath,amssymb,graphicx}
\usepackage{color}
\begin{document}

\title{Absorption-induced transparency metamaterials in the terahertz regime}

\author{Sergio G. Rodrigo}\email{Corresponding author: sergut@unizar.es}
\affiliation{Centro Universitario de la Defensa, Ctra. de Huesca s/n, E-50090 Zaragoza, Spain}
\affiliation{Instituto de Ciencia de Materiales de Arag\'{o}n and
Departamento de F\'{i}sica de la Materia Condensada, CSIC-Universidad de Zaragoza, E-50009, Zaragoza, Spain}  

\author{L. Mart\'in-Moreno}
\affiliation{Instituto de Ciencia de Materiales de Arag\'{o}n and
Departamento de F\'{i}sica de la Materia Condensada, CSIC-Universidad de Zaragoza, E-50009, Zaragoza, Spain}  

\begin{abstract}
Contrary to what might be expected, when an organic dye is sputtered onto an opaque holey metal film, transmission bands can be observed at the absorption energies of the molecules. This phenomenon, known as absorption-induced transparency, is aided by a strong modification of the propagation properties of light inside the holes when filled by the molecules. Despite having been initially observed in metallic structures in the optical regime, new routes for investigation and applications at different spectral regimes can be devised. Here, in order to illustrate the potential use of absorption induced transparency at terahertz, a method for molecular detection is presented, supported by a theoretical analysis. 
\end{abstract}

%\setboolean{displaycopyright}{true}

\maketitle
%\thispagestyle{fancy}
%\ifthenelse{\boolean{shortarticle}}{\abscontent}{}

%\section{Introduction}
Absorption-Induced Transparency (AIT) is an optical phenomenon that can be observed when optically active dye molecules are deposited on top of a metallic array of holes~\cite{HutchisonACI11}. Surprisingly, an initially opaque metallic hole array (HA) may become translucent with the incorporation of the molecular compound, at the spectral range where the molecules absorb electromagnetic radiation. This phenomenon results from two physical contributions~\cite{Zhong15}: one is due to the excitation of surface bounded waves when the molecules form a thin layer on top of the system, the other contribution originates from the modification of the propagation constant of light inside the holes when the molecules can penetrate into them. The AIT phenomenon was first observed in the visible range, but it was later suggested that AIT can operate in different spectral regimes~\cite{RodrigoPRB13}. 

%The prediction has been experimentally confirmed with the use of lithium fluoride as the filling medium, a solid compound with absorption lines at THz~\cite{MerinoJEurCeramSoc13,Acosta15}.

\begin{figure}[htbp]
\centering
\fbox{\includegraphics[width=0.6\linewidth]{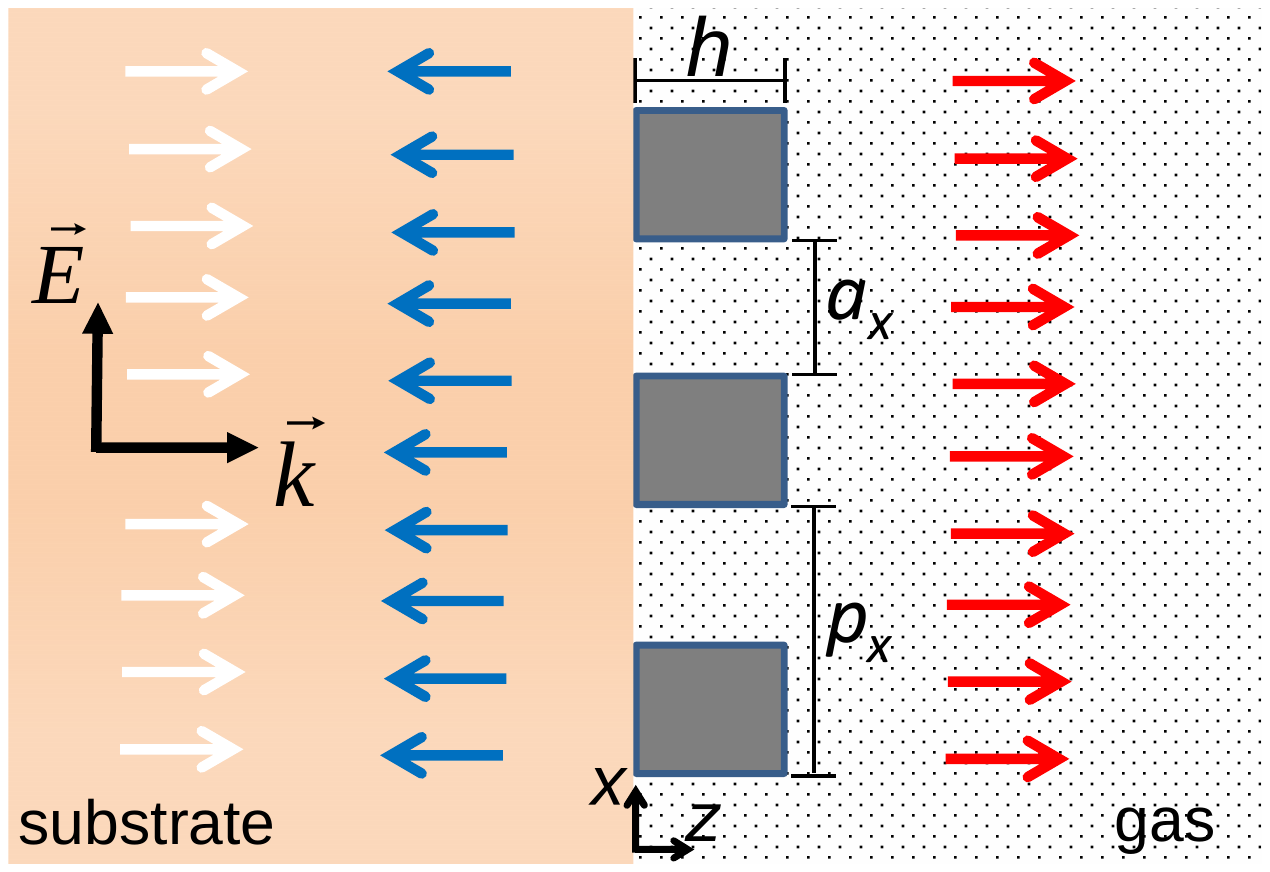}}
\caption{Principle of the AIT-based detection method. Transmitted (right) and reflected (left) THz radiation carries information about whether or not a given target substance is present inside the holes of a holey metal film.}
\label{fig:false-color} \label{fig0}
\end{figure}

Particularly interesting may be the application of AIT to the terahertz (THz) range, as this is one of the most promising spectral windows for its present and future technological applications~\cite{TonouchiNaturePhoton07,MittlemanNatPhoton13}. Many everyday materials (cardboard, plastics, ceramics...) are transparent at THz, and the THz waves are safe for its use in biological systems, so imaging and security screening stand out of the possible applications investigated so far~\cite{LiuIEEE07}. Detection of chemical compounds with spectral signatures at THz (like biomolecules, some drugs and explosives) is also a very fruitful field of research~\cite{NagelAPL02,ChenOptExp07,ParkNanoLett13}. 

One focus of research in the THz range are metamaterials, i.e., artificial composite systems that behave like an effective uniform medium with an optical response that is mainly controlled by the size and the shape of their geometrical features~\cite{ReinhardIEEE13}. Nano- and micro- structured metals additionally provide both enhancement and confinement of THz fields in small volumes, which enabled waveguiding~\cite{WilliamsAPL09} and more exotic applications like detection of microorganisms~\cite{ParkSciRep14}, among others.

In this letter we establish simple guidelines to achieve AIT in the THz regime. We theoretically demonstrate how the AIT phenomenon can be found at THz by analyzing an exemplary system. The paradigmatic AIT configuration is depicted in Fig.~\ref{fig0}. A metallic HA is placed on top of a dielectric substrate. A gaseous substance is released in the air region. Conversely to those approaches designed to feature strong resonances at the absorption energies of the molecules~\cite{ParkNanoLett13} and, with typical functionalities based on accurately measuring spectral shifts in the presence of a given substance~\cite{OHaraJIMTWaves12}, our proposal relies on the second mechanism behind AIT briefly discussed in the introduction; the appearance of a spectral feature in transmission and reflection induced by the molecular specimen when this fills the holes~\cite{RodrigoPRB13}. 

For proof of principle, we consider the following materials: the holey film is metallic, which we treat as a perfect electrical conductor (PEC, $\varepsilon_{metal}=-\infty$), this being a good approximation in the THz range for metals such as gold, silver, aluminum or copper. The HA is placed on top of a transparent polytetrafluoroethylene (PTFE or Teflon) substrate, $\varepsilon_{\text{PTFE}}=2.06$~\cite{HejaseIEEE11}. We choose the poisonous gas hydrogen cyanide (HCN) as the substance to be detected in the air region. Hydrogen cyanide is exhaled by motor vehicles, which is also present in tobacco in low concentrations and is extensively used in industry and manufacturing. This gas can kill a person for concentrations $\geq $300~ppm, in a few minutes. One of its multiple absorption resonances at THz is a narrow line spectrally located at $\lambda_A \approx 1239.89$~GHz, with a full-width at half-maximum (FWHW) $\approx 126$~MHz~\cite{BigourdOptLett06}. To characterize AIT at THz we will focus on that feature throughout this work. The contribution to the dielectric constant of HCN from this absorption line is well represented by a Lorentz function in the frequency domain, which provides suitable  optical response in the spectral window of interest for this gas, $\varepsilon_{\text{HCN}}(\nu)=\varepsilon_{inf} - \Delta \varepsilon \, \Omega^2/(\nu^2-\Omega^2+\imath \nu \Gamma)$. We obtained $\varepsilon_{inf}=1.0$, $\Delta \varepsilon=3.5 \times 10^{-9}$, $\Gamma=96.79$~GHz, and  $\Omega=1239.89$~GHz for a HCN concentration of $\approx 200$~ppm. We used these parameters to reproduce the transmission curve reported in Fig.~3 of Ref.~\cite{BigourdOptLett06}. Additional absorption resonances could be easily incorporated into the model by adding extra Lorentzian terms. 

We design the geometry in order to obtain a maximum contrast in transmission and reflection between filled and empty holes within the AIT spectral band. To achieve this objective the bare HA (system without molecules) must reflect almost all THz light back in the spectral window of interest, around the HCN absorption line. For the sake of simplicity, the THz source illuminates the system at normal incidence and is polarized so that the electric field oscillates parallel to the x-axis. Note however that the AIT band is almost flat within a wide range of incidence angles~\cite{Zhong15}. With this in mind, the design of an AIT configuration for detection purposes is quite simple. Once the absorption line of a target molecule for screening, $\lambda_{A}$, has been selected, AIT essentially depends on two geometrical parameters: the spatial period and the dimensions of the holes. The bare HA must behave as a mirror, so it is necessary to avoid any kind of enhanced optical transmission phenomena and lattice effects like Wood's anomalies. It is well known that the excitation of surface bounded waves can enhance transmission in arrays of subwavelength holes~\cite{EbbesenNature98,MartinMorenoPRL01,LiuNature08}. The spectral locations of the peaks generated are found relatively close to the Wood's anomalies. Therefore $\lambda_{A} > n_{\text{PTFE}} \, p_{y}$, the right-hand of this inequality closely indicates the largest wavelength these lattice effects can be found, given the following choice of periods $p_{y} > p_{x}$ and substrate $n_{\text{PTFE}} > n_{air}=1$. Localized resonances~\cite{KoerkampPRL04,DegironOptCommun04} must also be avoided in the AIT-based metamaterial. These resonances, occurring close to the cutoff frequency of the hole, can be seen either as zero-order Fabry-Perot resonances~\cite{GarciaVidalPRL05,CarreteroPalaciosPRB12} or transmission in an epsilon-near-zero material~\cite{SilveirinhaPRL06}. The cutoff wavelength of an empty hole, $\lambda_c$, has to appear slightly blueshifted regarding $\lambda_{A}$. Given that $\lambda_c= 2\,a_y$ for an unfilled rectangular waveguide in PEC (we choose $a_y > a_x$), it is the long side of the rectangle which has to be properly adjusted. The geometrical parameters we have chosen to fulfill all these requirements consist of rectangular holes with size $a_x=50.0\,\mu \text{m}$ and $a_y=120.9\,\mu \text{m}$, arranged in a rectangular lattice with $p_x=80\,\mu \text{m}$ and $p_y=150\,\mu \text{m}$, milled in a thick metal film ($h=1.0\,\text{mm}$). We have check that the device performance is not impaired even if the film thickness, periodicity and short side of the rectangles are changed by several microns. The long side of the rectangles would require more precise tunning during the fabrication process, with a tolerance range of approximately $\pm 500$~nm. This accuracy is perfectly available with current nanofabrication techniques~\cite{ChenNatCommun13} that allowed to produce gaps as small as $2$~nm in micro sized structures~\cite{ParkFaradayDiscuss15}. Although a regular lattice has been chosen because it is a robust, easier to fabricate at large scale system and that allows tight packaging, note however that AIT is determined by the guiding properties of light of a single hole when filled by the molecules, so AIT can be found in isolated holes~\cite{RodrigoPRB13,Zhong15}. Therefore, the use of different hole shapes~\cite{LockyearPRL05,OrbonsOptExpress06,RodrigoOptExp10} and spatial distributions of holes would provide of additional freedom to exploit AIT.

Our theoretical analysis is based on the Coupled Mode Method (CMM), a numerical tool which has been successfully applied to study the optical response of holey metal films~\cite{MartinMorenoJoP08}. Maxwell's equations are solved by expanding the electromagnetic (EM) fields in the different regions of space, transmission and reflection coefficients are thus calculated by imposing appropriate boundary conditions (for further details see Ref.~\cite{GarciaVidalRevModPhys09}). 

Metallic HAs can be described as homogeneous film metamaterials under the effective medium limit ($\lambda >> p_{y}$, in our case), where diffraction effects can be neglected~\cite{SmithPRB02} and each hole plays the role of a metaatom. Following a homogenization procedure, which it is also based on the CMM method~\cite{PendryScience04,MaryPRB09}, the effective parameters of the homogeneous film are found: 
\begin{equation}
\varepsilon_{eff}=(\varepsilon_{\text{hole}}-(\lambda/\lambda_c)^2)/S^2\label{eq1}
\end{equation}
for the dielectric constant and $\mu_{eff}=S^2$ for the effective magnetic susceptibility. Here $S=\frac{2\sqrt{2}}{\pi}\sqrt{\frac{a_x a_y}{p_x p_y}}$. In this model the refractive index, $n_{eff}=(\varepsilon_{eff} \mu_{eff})^{1/2}$, is fixed by the propagation constant of a single hole as follows; $k_{\text{hole}}=n_{eff} \, k_0$, where $k_0=2 \pi / \lambda$ being $\lambda$ the wavelength of the incident light. The effective medium approach suffices to describe the optical properties of the system investigated, as shown later on.

\begin{figure}[htbp]
\centering
\fbox{\includegraphics[width=0.45\linewidth]{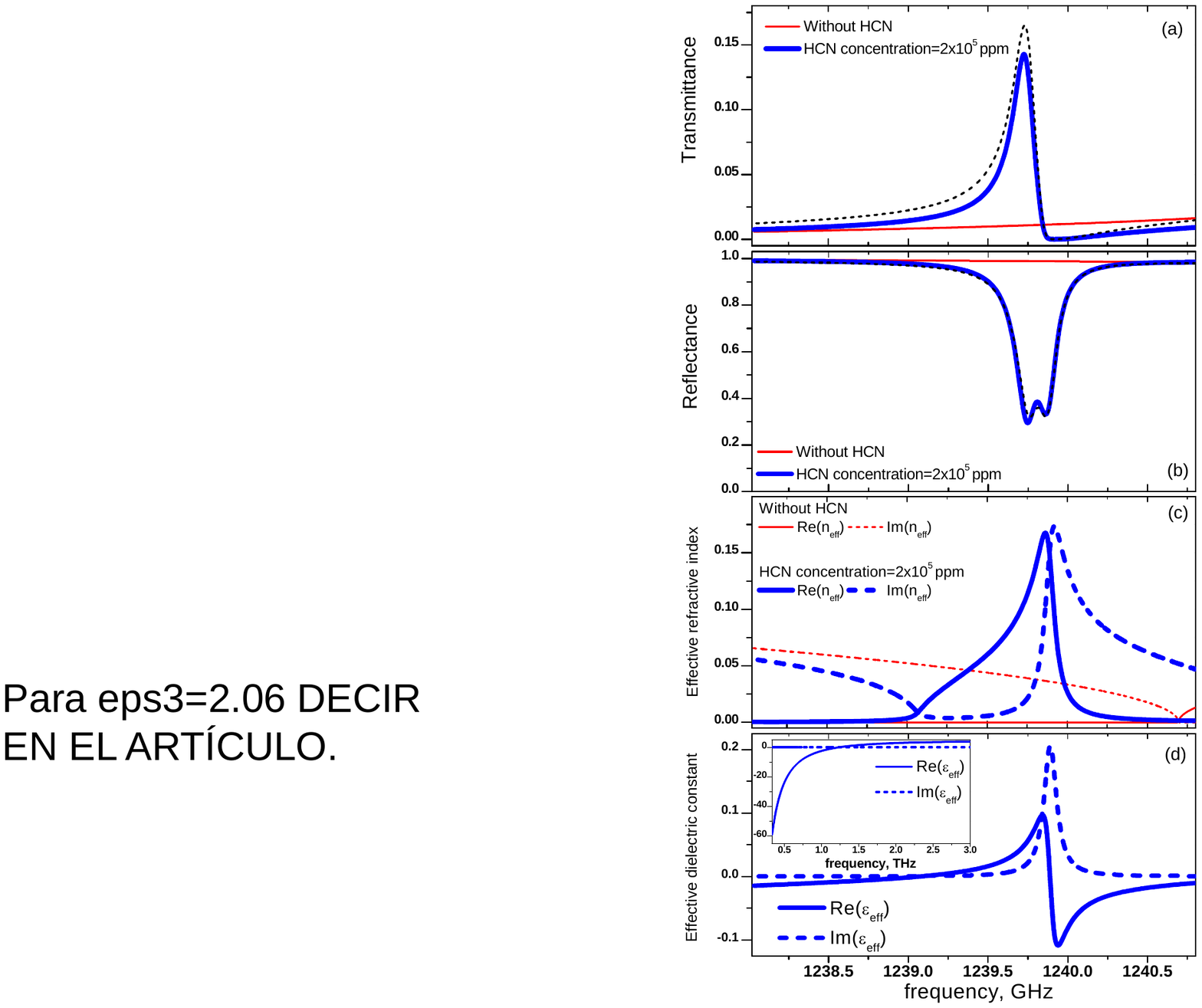}}
\caption{(a) and (b) Transmission and reflection spectra calculated with the CMM method in the absence (red thin curves) and the presence (blue thick curves) of HCN inside the holes (air outside). The hole array is made of rectangular holes ($a_x=50.0\,\mu \text{m}, a_y=120.9\,\mu \text{m}$) arranged in a rectangular lattice ($p_x=80\,\mu \text{m}$, $p_y=150\,\mu \text{m}$) and milled in a $1.0 \, \text{mm}$ thick metal film, everything on top of a transparent PTFE (Teflon) substrate, $\varepsilon_{PTFE}=2.06$~\cite{HejaseIEEE11}. The dashed lines shows transmission and reflection for a uniform layer characterized by $\mu_{eff}$ (see main text, for a definition) and $\varepsilon_{eff}$ [panel (d)]. The THz source is aligned so the system is illuminated at normal incidence. The electric field points along the x-direction [Fig.~\ref{fig0}]. (c) The imaginary (real) part of the effective refractive index is shown with dashed (solid) lines in the absence (red thin curves) and the presence (blue thick curves) of HCN inside the holes. (d) The corresponding effective dielectric constant of such metamaterial, described by Eq.~\ref{eq1}.}
\label{fig:false-color} \label{fig1}
\end{figure}

In Fig.~\ref{fig1}(a)-(b) transmission and reflection spectra are shown with thin red lines for a system without HCN. Clearly the prerequisite of low transmittance ($T\approx 0$) and high reflectance ($R\approx 1$) is achieved for the chosen parameters. To detect transmission when the gas is present in the system, low concentrations in the output region are mandatory as otherwise no significant signal would reach the detector due to attenuation. Assuming the HCN gas is filling the holes for $2\times 10^5$~ppm concentration (air outside) a AIT feature shows up in transmittance [Fig.~\ref{fig1}(a), thick line] and a double dip spectral feature is seen in reflection [Fig.~\ref{fig1}(b), thick line], in accordance with previous predictions~\cite{RodrigoPRB13}.

The effective dielectric constant is shown in the main panel of Fig.~\ref{fig1}(d). It is clearly dominated by a bounded charge-like response (Lorentzian function) centered at the absorption line of HCN. The inset shows the same but spanning the whole THz range. A Drude-like behavior of geometrical origin dominates in the large frequency scale.  The last interpretation is readily confirmed analyzing Eq.~\ref{eq1} and noting that $\varepsilon_{\text{hole}}=\varepsilon_{\text{HCN}}$ for the AIT configuration. For this system the validity of the effective medium approach is confirmed in Fig.~\ref{fig1}(a)-(b), which shows the good agreement between transmission and reflection calculated for both a uniform layer characterized by $\mu_{eff}$ and $\varepsilon_{eff}$ (shown with dashed lines) and the full calculations (blue thick lines). 

\begin{figure}[htbp]
\centering
\fbox{\includegraphics[width=0.6\linewidth]{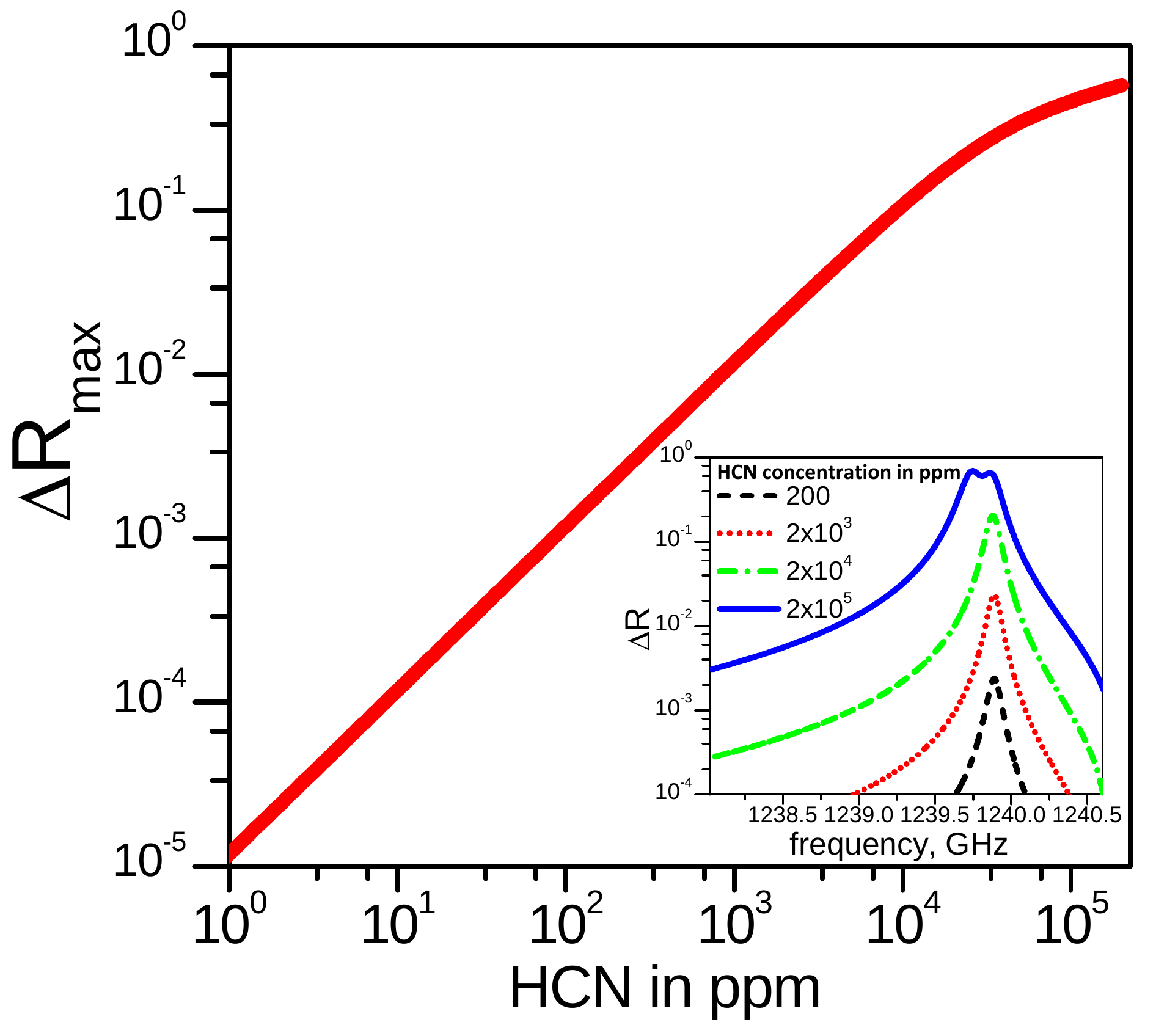}}
\caption{Reflection contrast, $\Delta R$, as a function of the amount of HCN within the holes (in ppm) at the spectral locations of $\Delta R$ maxima. Inset: $\Delta R$ as function of the frequency for different HCN concentrations. Same system of Fig.~\ref{fig1} (for the rest of definitions, see main text).}
\label{fig:false-color} \label{fig2}
\end{figure}

Equation~\ref{eq1} shows that the bare HA behaves as an ordinary metal for wavelengths larger than $\lambda_c= 2\,a_y$ (frequencies shorter than $\sim 1240.69$~GHz), which explains its optical response. 
%In Fig.~\ref{fig1}(c), the corresponding $n_{eff}$ is represented with thin lines. In the metallic region  $Re(n_{eff})=0$, so the effective wavelength of light for such metamaterial, $\lambda_{eff} =\lambda/n_{eff}$, tends to infinity.

For filled holes, the AIT spectral position, intensity and spectral width is mainly controlled by $n_{eff}$~\cite{RodrigoPRB13}.
There are spectral regions where the influence of the molecules is weak and $Re(n_{eff})\approx 0$ [Fig.~\ref{fig1}(c), thick solid line], like for the bare HA when it behaves as a metal [Fig.~\ref{fig1}(c), thin solid line]. In these regions the effective refractive index is almost a purely imaginary complex number, which provokes a large impedance mismatch at the metamaterial boundaries so $T \approx 0$ and $R\approx 1$.  Around the HCN absorption frequency, $Re(n_{eff})$ becomes different from zero making light transmission possible within the normal dispersion region, given the moderate values of $Im(n_{eff})$ [Fig.~\ref{fig1}(c), thick dashed line]. In contrast, both the transmission minimum [Fig.~\ref{fig1}(a)] and the highest frequency dip in reflection [Fig.~\ref{fig1}(b)] occur within the anomalous dispersion region of $n_{eff}$. In this spectral window $Im(n_{eff})$ is so high  that light is almost completely absorbed. 

The AIT configuration used for detection could operate in transmission or reflection mode (or both) depending on the setup conditions. One have to be sure that the gas is filling the holes in both cases, at least partially~\cite{RodrigoPRB13}. For gas detection in transmission mode the output region should be free of gas as explained before. We have checked that reflectance does not change appreciably if HCN is everywhere in the transmission region, which is an advantage of working in reflection mode. The reflection mode has the additional advantage that the dielectric substrate acts as a shield to prevent HCN leakage from the transmission region to the reflection region, which are thus isolated from each other. In that case, both detector and source could be safely located at the same side, with the molecules dispersed in the transmission zone (a reaction chamber, for instance). The detection of the molecules relies then on the reflection contrast, defined as: $\Delta R=R_{\text{HA}}- R_{\text{HA+HCN}} \approx 1-R_{\text{HA+HCN}}$. In the inset of Fig.~\ref{fig2}, $\Delta R$ is represented as a function of frequency, for different HCN molecular concentrations. In the main panel, the maximum value $\Delta R$ is shown as a function of the density of HCN in the system. In order to exploit the AIT configuration at fixed wavelength and low concentrations, a potential experimental setup should incorporate high sensitivity and spectral resolution detectors~\cite{CriadoIEEE13}, able to discriminate small reflection changes at fixed wavelengths. A contrast value of $\Delta R_{\text{max}} \approx 10^{-4}$ is necessary to detect at the 9~ppm level (see Fig.~\ref{fig2}), which for HCN has been considered competitive for THz spectroscopy as compared to chemical analytical methods~\cite{BigourdApplPhysB07}.

In conclusion, we have introduced the use of AIT-based metamaterials for applications in the THz range by presenting a detection method. Exemplarily, we have proposed an AIT system designed to recognize the presence of the poisonous HCN gas. Nevertheless, the proposed method is neither constrained to a specific absorber nor to the geometrical parameters used. As a component of a sensing device, the AIT-configuration would provide field enhancement, introducing further advantages over bulk THz spectroscopy, like chemical functionalization for molecular trapping. The pattering of different holes (periodically arranged or not) with different shapes and sizes on the same sample, each one designed for covering one of several characteristic wavelengths of a single molecular compound (or a mixture of them) might lead to multi-spectral detection based on AIT. Moreover, the AIT configuration could operate as a spectrally tunable metasurface at THz. By introducing inside the holes optically and/or electronically driven natural materials (like gallium arsenide) transmission and reflection could be ``dialed'' for THz radiation filtering. Therefore, the AIT configuration might become a complementary method to existing ones for applications in sensing, light filtering, enhanced absorption and polarization optics at THz~\cite{OHaraJIMTWaves12}.  More speculative, AIT-based systems could be exploited in THz field-matter interaction research led by recent advances in intense THz sources~\cite{GuNatCommun12,KampfrathNaturePhoton13}. 

\textbf{Acknowledgment.} The authors thank the funding of the Spanish Ministry of Economy and Competitiveness under project MAT2014-53432-C5-1-R.
%\textbf{Acknowledgment.} The authors thank H. Haase, C. Wiede, and J. Gabler for technical support.

%\section*{References}

%\bigskip
%\noindent Add citations manually or use BibTeX. See \cite{Zhang:14}.

% Bibliography
%\bibliography{TesisSGR18ene10}

\end{document}